\documentclass[preprint]{aastex}
\usepackage{rotate}
\usepackage{color}
\usepackage{graphicx}
\shorttitle{Thermal absorption as the cause of GPS}
\shortauthors{Lewandowski et al.}
\begin{document}

\def\ni{\noindent}
\def\be{\begin{equation}}
\def\ee{\end{equation}}
\def\lesssim{\raisebox{-0.3ex}{\mbox{$\stackrel{<}{_\sim} \,$}}}
\def\gtrsim{\raisebox{-0.3ex}{\mbox{$\stackrel{>}{_\sim} \,$}}}

\title{Thermal absorption as the cause of gigahertz-peaked spectra in pulsars and magnetars}

\author{Wojciech Lewandowski,\altaffilmark{1}}
\author{Karolina Ro{\.z}ko,\altaffilmark{1}}
\author{Jaros{\l}aw Kijak,\altaffilmark{1}}
\author{George I. Melikidze\altaffilmark{1,}\altaffilmark{2}}

\affil{\altaffilmark{1}Kepler Institute of Astronomy, University of Zielona Gora, Lubuska 2, 65-265
Zielona G\'ora, Poland}
\affil{\altaffilmark{2}Abastumani Astrophysical Observatory, Ilia State
University, 3-5 Cholokashvili Ave., Tbilisi, 0160, Georgia}

\email{boe@astro.ia.uz.zgora.pl}

\begin{abstract}
We present a model that explains the observed deviation of the spectra of some pulsars and
magnetars from the power-law spectra which are seen in the bulk of the pulsar population. 
Our model is based on the assumption that the
observed variety of pulsar spectra can be naturally explained by the thermal free-free absorption
that takes place in the surroundings of the pulsars. In this context, the variety of the pulsar spectra can be explained according to the shape, density and temperature of the absorbing media and the optical path of the line-of-sight across that. We have put specific emphasis on the case of the radio magnetar SGR J1745-2900 (also known as Sgr A* magnetar), modeling the rapid variations of the pulsar spectrum after the outburst of Apr 2013 as due to the free-free absorption of the radio emission in the electron material ejected during
the magnetar outburst. The ejecta expands with time and consequently the absorption rate decreases
and the shape of the spectrum changes in such a way that the peak frequency shifts towards the
lower radio frequencies. In the hypothesis of an absorbing medium, we also discuss the similarity
between the spectral behaviour of the binary
pulsar B1259$-$63 and the spectral peculiarities of isolated pulsars. 
\end{abstract}
\keywords{pulsars: general --- }

\section{Introduction}

The gigahertz-peaked spectra (GPS) pulsars are those radio pulsars that show a broad maximum of the flux density at frequencies typically around one or few GHz. This   significantly distinguishes them from
the bulk of the pulsar population. The radio spectra of pulsars can be generally described by a single power-law function, with average spectral index which are in the range from about $-1.4$ to about $-1.8$, according to different analyses and investigated samples \citep[e.g. ][]{lorimer95,kramer98,Maron2000,Bates13}. The distribution is also rather broad, with a standard deviation around 1 when a gaussian distribution for the indexes is adopted (Bates at al. 2013).

The first pulsars that indicate a turnover at high frequencies were identified by \citet{Kijak07},
and later \citet{Kijak11a} named them ``GPS pulsars''. \citet{Kijak11a} also indicated that the GPS
pulsars tend to have peculiar environments, such as pulsar wind nebulae (PWN), dense H~II regions
or supernova remnants (SNR). Therefore, it was suggested that the external influence accounts for
the GPS phenomena rather than the peculiarities of the emission process. This statement was further
strengthened by \citet{Kijak11b} who demonstrated that the spectrum of the binary pulsar B1259$-$63
varies according to the pulsar's orbital phase.

Later, based on the available data \citet{Kijak13} showed that two radio-magnetars (J1550$-$5418
and J1622$-$4950) can also be considered as a GPS sources, since they show the turnovers at
frequencies of about a few gigahertz.  The number of known GPS pulsars is still growing and
currently there are eleven such objects \citep{Dembska14a} which seem to be a rather small
population. However, recent statistical studies of the largest pulsar search surveys
\citep{Bates13} indicate that the GPS pulsars can amount up to several percent of the whole pulsar
population. Thus we can expect to observe up to 200 pulsars showing the GPS phenomenon among over
2300 currently known pulsars. The reason why we currently observe only 11 GPS pulsars should rather
be an observational selection effect. The spectra of GPS pulsars make them easier to be
discovered at intermediate frequencies (around 1.4 GHz). But not all discoveries are followed by
observations at lower frequencies, therefore, the spectra of newly discovered pulsars usually
remain unknown, which partially is caused by the relatively poor sensitivity (at least until
recently) of the low frequency observations. In addition, the GPS pulsars, by definition, are
weaker at sub-gigahertz frequencies. Fortunately the situation has recently changed as a number of
low frequency observatories (LOFAR, MWA, LWA) are becoming operational.

As we have already mentioned, the case of the binary pulsar B1259$-$63 is the key to understanding
the origin of the GPS phenomenon in pulsars. For this case \citet{Kijak11b} suggested that the
primary reason of the spectral turnover is the thermal free-free absorption of the pulsar radio
emission in the strong stellar wind of the companion Be-star. Therefore, the pulsar motion along
the elliptical orbit causes spectral evolution. While the pulsar is staying far from the companion
(where the stellar wind density is quite low) its spectrum is a usual, single power-law. However,
when the pulsar is approaching periastron passing through the dense stellar wind its spectrum
displays the GPS behavior. This effect has been recently modeled by \citet{Dembska14b}, who have
shown that the evolution of the spectrum can be explained by free-free absorption in the stellar
wind. If the wind carries about $10^{-9} M_{\sun}$~year$^{-1}$  the mass loss yields the electron
density of about $10^5$ electrons per cm$^3$ at the distance of 1~AU from the star. This density is
enough to provide the observed absorption. Let us note that, the pulsar approaches the star as
close as 0.4~AU in the periastron.

\citet{Sieber1973} was the first who proposed that the thermal absorption in the interstellar medium
caused low-frequency turnovers in pulsar spectra. But it should be mentioned that the Vela pulsar
having the highest turnover frequency (with peak at about 600~MHz) quoted in that paper is
currently known as having a bow-shock PWN.

In the case of PSR~B1259$-$63 the apparent cause of turnovers is the absorption in the dense
stellar wind, while it is not the case for other GPS pulsars as they do not have stellar
companions. However, as it was mentioned above the GPS pulsars tend to adjoin some peculiar
environments where the matter density is much higher than that of the ISM. Thus the natural
question arises, can the same process be responsible for the GPS phenomenon in all objects under
consideration? We can distinguish three cases: first of all the case of PSR~B1259$-$63 which is
expected to have a highly localized (within a few A.U. inside the orbit) environment with a high
particle density and a high temperature (presumably of the order of several thousand kelvins);
secondly, there is a vast interstellar space of very low electron density, which is relatively cold
and causes low-frequency turnovers in pulsars; and finally, there are these peculiar pulsar
surroundings, which are usually much larger in size than PSR~B1259$-$63 orbit, but much smaller and
at the same time much denser than the interstellar free-electron clouds.

In this paper we explore the possibility whether the thermal absorption in pulsar surroundings can
indeed cause high-frequency turnovers in the GPS pulsars, as we proposed in
\citet{Kijak11b} and \citet{Kijak13}. Using the simplified model of such
environments we estimate the domain of the parameters that are necessary to cause the thermal
absorption of the pulsar radiation, as well as to change the pulsar spectrum in such a way that it
becomes a GPS. The main parameters to be estimated are the size of the absorbing regions, the
density and temperature of electrons. Since at present it is extremely hard to measure these
parameters (e.g. the electron temperature and distribution within PWNe or SNRs), we hope that our
simulations at least give a rough estimate of what we expect, or rather what we need in order to
explain the GPS pulsar spectra using the thermal free-free absorption.

\section{Thermal absorption as the cause of GPS}

In our model we consider the propagation of pulsar radio emission through the absorbing region \citep{Kijak11b,Kijak13}. We
assume that the intrinsic pulsar spectrum is a single power-law with a spectral index equal or
close to $-1.8$ (which is an average population value, \citealt{Maron2000}). In our models we use
the approximation of homogeneous electron density and uniform temperature since we need only rough
estimations of the parameters. We assume that the absorption happens over a fraction of the
pulsar's line-of-sight and the optical path length is defined by the geometry.

\subsection{The model}

The model is based on the formalism of the radiative transfer theory. For the simplicity we also
assume the local thermal equilibrium approximation. Under such conditions the radiative transfer
equation has the following solution \citep{Rohlfs04}:
\begin{equation}\label{rad_tra}
I_{\nu}(s) = I_{\nu}(0) e^{-\tau_{\nu}(s)} + B_{\nu}(T) (1-e^{-\tau_{\nu}(s)}),
\end{equation}
where $\tau_{\nu}$ is the optical depth, $B_{\nu}(T)$ is the Planck function and $I_{\nu}(0)$ is the incoming
intensity (which in our case is the intrinsic emitted intensity). Assuming that the
plasma is quasi-neutral and the electrons obey the thermal distribution we obtain the optical depth
in the following form \citep{Rohlfs04}:
\begin{equation}\label{opt_dep}
\tau_{\nu} = 3.014 \times 10^{-2} \left( \frac{T_{\mathrm{e}}}{\mathrm{K}} \right)^{-3/2}
\left( \frac{\nu}{\mathrm{GHz}} \right)^{-2} \left( \frac{\mathrm{\mathrm{EM}}}
{\mathrm{pc}~\mathrm{cm}^{-6}} \right) <g_{\mathrm{ff}}>,
\end{equation}
\noindent
where $EM$ is the Emission Measure ($EM = \int N_e^2 dl$) and $<g_{\mathrm{ff}}>$ is the Gaunt factor \citep[see][for
definitions]{Rohlfs04}. In the case of the free-free absorption and an uniform density profile the
$EM = N_e^2 d$, where $d$ is simply the total thickness of the absorber.

Equations [\ref{rad_tra}] and [\ref{opt_dep}] as well as the geometrical length of the path which
the radio waves follow through the absorbing region allows us to estimate the fraction of the flux
that is absorbed during its passage through the medium. The value of the fraction depends on the
physical parameters of the absorbing region, as well as on the wave frequency, i.e. the amount of
the absorbed radiation obviously depends on the optical depth which on its turn is frequency
dependant. Applying results of the calculations of optical depth to the simulated (intrinsic)
pulsar spectra we can estimate the shape of ``observed" spectra.

\begin{figure}
\includegraphics[angle=-90,width=8cm]{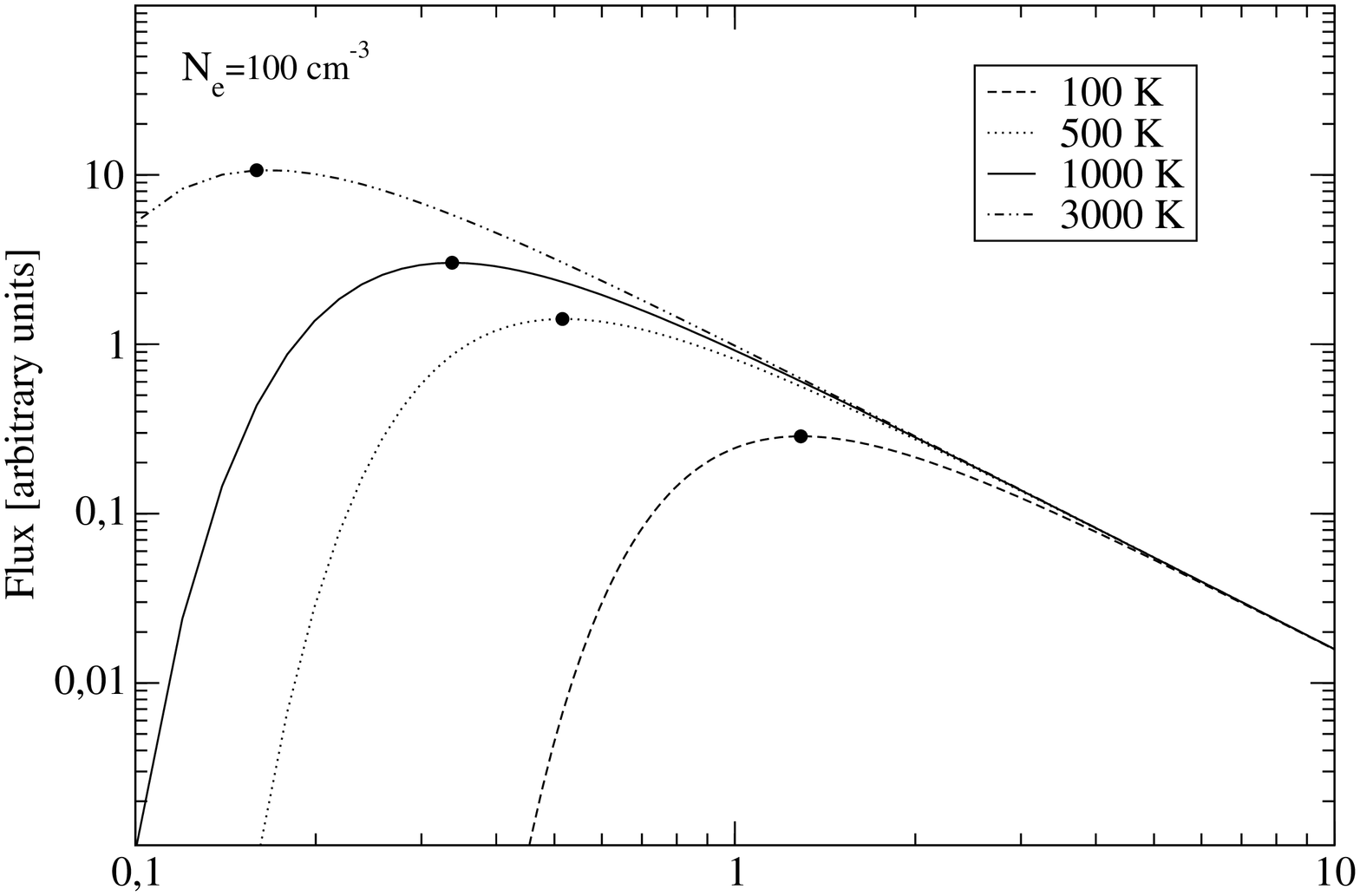}
\includegraphics[angle=-90,width=8cm]{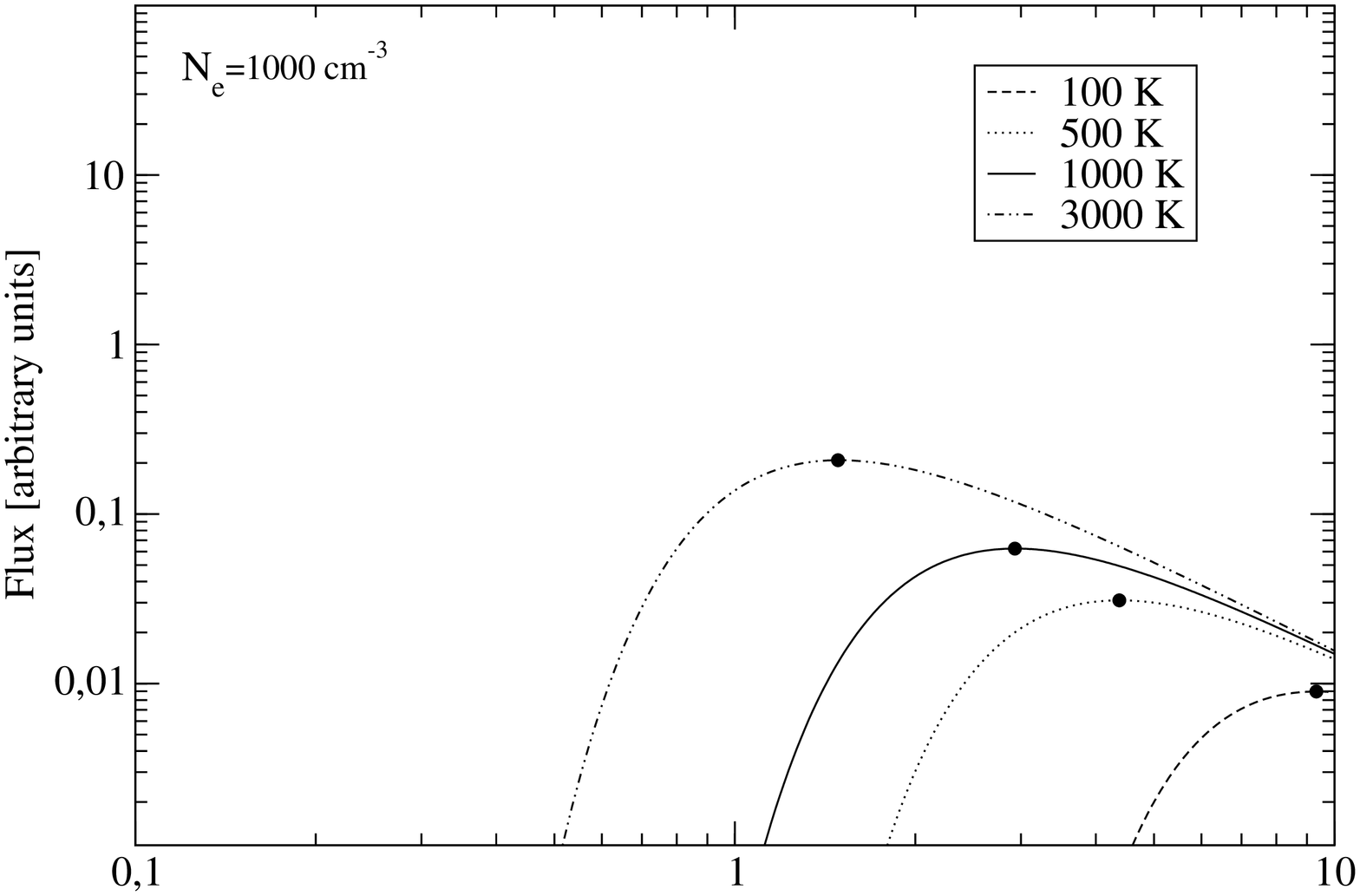}
\includegraphics[angle=-90,width=8cm]{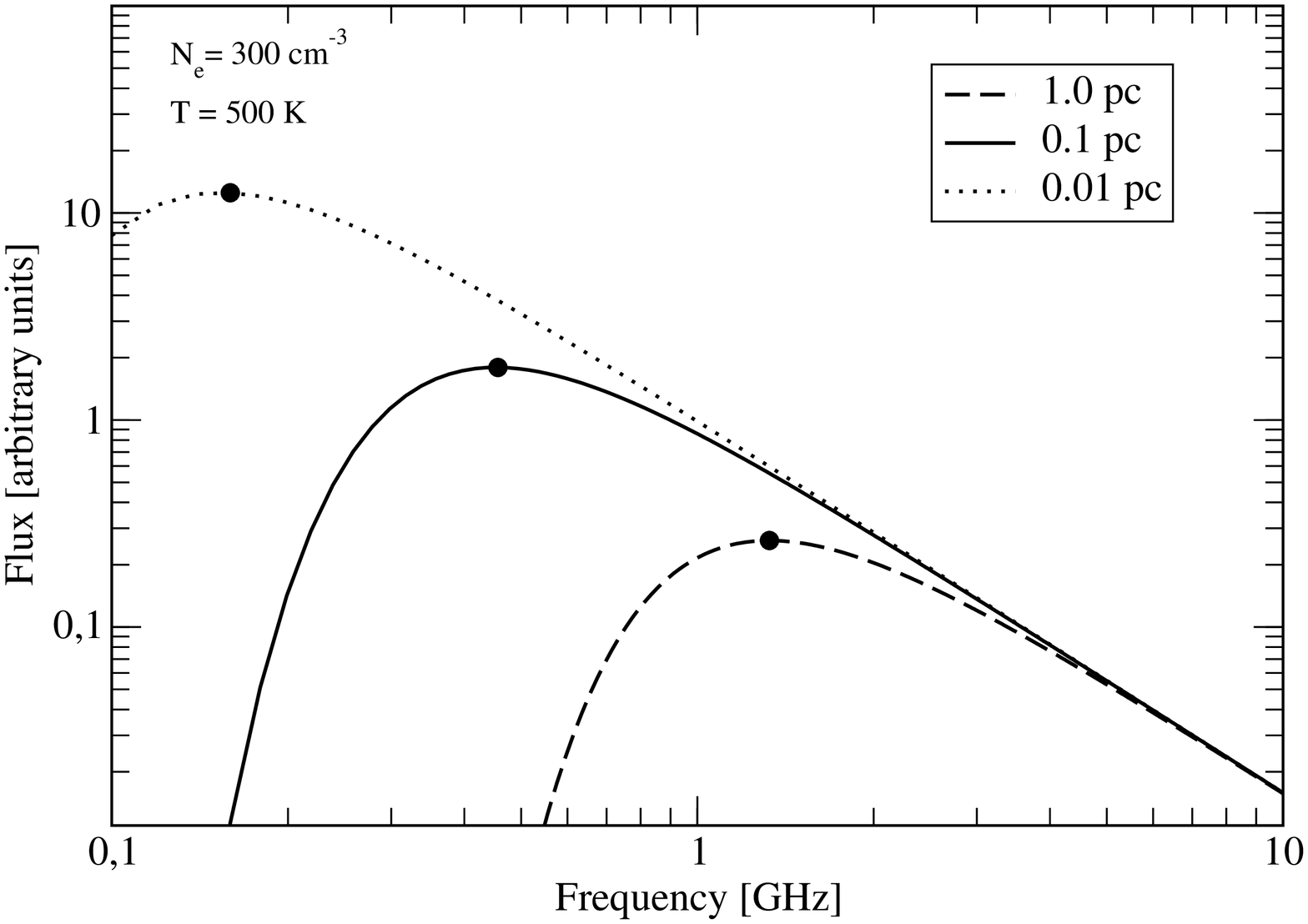}
\caption{
Simulated spectra of a pulsar in the case of the presence of an absorber with thickness of 1.2 pc.
The top panel shows the spectra for a case where the electron density is equal to
$100~\mathrm{cm}^{-3}$ and different temperatures of the absorbing electrons. The middle panel
shows spectra simulated for the same temperatures of the absorber but the electron density was set
to $1000~\mathrm{cm}^{-3}$. The bottom panel shows the simulated pulsar spectra for various
thicknesses of the absorbing region. In this case the electron density and the electron temperature
are assumed to be equal to $300~\mathrm{cm}^{-3}$ and 500~K, respectively.
\label{fig1} }
\end{figure}

The simplest possible case to be considered is when the cloud of an absorbing electron-rich
material is located anywhere between the pulsar and the observer. The cloud does not have to be
close to the pulsar as in our calculations the position of the absorbing region along the
line-of-sight does not matter. However, we are mostly interested in the cases where a cloud is
located in the vicinity of a pulsar, thus the pulsars are somehow physically connected to the
absorbers (e.g. PWNe, SNRs or some other regions with heightened density).

The top and middle panels of Figure~\ref{fig1} show the simulated pulsar spectrum for the absorber
whose thickness along the line-of-sight is 1.2~pc. The panels show the spectra obtained for various
values of the electron temperature with the assumption that the electron number density is constant
and equals to 100 cm$^{-3}$ for the left panel, while for the right panel the electron number
density equals to 1000 cm$^{-3}$. As it can be seen on the panels the amount of absorption
obviously increases with decreasing the temperature. For the lower density the temperature should
be considerably low (of the order of 100~K) to be able to affect the pulsar spectrum significantly
and produce a turnover at gigahertz frequencies. At higher temperatures the turnover frequency
decreases below 1~GHz, approaching 100~MHz at temperatures of the order of a few thousand kelvins.
Of course, the higher density translates into the higher optical depth (larger absorption) since in
the case of the free-free absorption the depth depends on $N_e^2$. In this case all the simulated
spectra show high frequency turnovers (above 1~GHz) with the spectrum peaking at frequencies close
to 10~GHz for the coldest case (100~K).

The bottom panel of Figure~\ref{fig1} shows the simulated pulsar spectra for three different values
of the geometrical length of the path (i.e. various thicknesses of absorbers) obtained for fixed
values of both the electron density and the temperature. In this case the values of the electron
number density and the temperature have been set to 300 cm$^{-3}$  and 500~K, respectively. One can
see that under such physical conditions one needs an absorber of about a parsec width to be able to
produce absorption which produces the pulsar spectrum peaking at frequencies close to 1~GHz.

Our calculations have demonstrated that in order to explain the observed GPS of pulsars one needs
an absorber of at least a fraction of a parsec thick, with relatively low temperature (preferably
below 1000~K) and with a significant electron number density of the order of a few hundred
particles per cm$^3$. The exact effect of the thermal absorption on the pulsar spectrum obviously
depends on the combination of all these three factors.

\subsection{SNR filaments}

Naturally a question arises whether in pulsar environments we can find some possible absorbers that
can fulfill the conditions implicated by our simulations. One of the main problems is connected
with the thickness of the absorbing electron-rich material as it is hard to imagine any homogenous
cloud of a size of a parsec within a supernova remnant.

When it comes to the observed densities and  the electron temperatures the available data are very
limited as it is not easy to measure these quantities. According to \citet{Graham1989} estimations
in the Crab Nebula filaments the electron density is about $200$ cm$^{-3}$.

The free electrons are partially created by the photo-ionization of hydrogen atoms in the Nebula
and it is easier to measure the number density of hydrogen than that of electrons. Using the
observations of ionized nitrogen and oxygen ([N~II] and [O~III]) \citet{Sankrit1998} estimated that
the hydrogen density in the filaments ranged from $1050$ to $2218$ cm$^{-3}$ and its temperature
was very high (over 8000~K). However, it should be mentioned that the Crab Nebula is quite a
special case and it is difficult to generalize the results to another SNRs.

Much higher densities were observed in very dense filaments of the G11.1-03 supernova remnant where
the estimated electron density was even as high as $N_{\mathrm{e}} = 6600
\pm 900~\mathrm{cm}^{-3}$, the temperature was about 5000~K and the thickness of the filament
was approximately 0.24 pc \cite{Koo07} .

It follows from all of the above that the filaments in the G11.1-03 SNR are the most promising
candidates for the source of the GPS phenomenon. In these filaments the free electron densities are
even much higher than the one we have used while simulating the spectra shown in the right panel of
Fig.~\ref{fig1}. Even if we assumed a slightly higher electron temperature (than what we used, see
Fig.~\ref{fig1}) the filaments would still provide enough absorption to obtain the spectral peak at
the gigahertz frequencies. In fact, for this particular filament (\citet{Koo07}) the simulated
pulsar spectra peaks at the frequency of 3~GHz assuming 6600~cm$^{-3}$, $5000$~K and $0.24$ pc for
the electron density, the electron temperature and the thickness, respectively.

The fact that only absorption in the SNR filaments can lead to the appearance of the GPS pulsars
naturally explains why most of the SNR-associated pulsars do not show this type of spectrum. It is
an observational selection effect since such filaments do not cover a nebula completely, the
line-of-sights of a very small fraction of pulsars pass through such filaments and, therefore, only
these pulsars show the GPS phenomena.

\subsection{Isolated GPS pulsars with PWNs}

In this section we explore the thermal absorption of the radio waves emitted from the pulsar in its
PWN. For our purposes we mostly consider the bow-shocks PWNe. There are several reasons why other
types of pulsar nebulae can be excluded. For example, the young PWNe (such as the Crab PWN) should
be excluded due to very high electron temperatures. As for the spherically symmetric PWNe they tend to
have a very low thickness.

The shape of the PWN depends mainly on its evolutionary stage. The details of the PWN evolution
theory can be found in \citet{Gaensler06}. Let us summarize it as follows. At the beginning the
shape of practically all PWNe is nearly spherically symmetric and the pulsar is located in the
centre of its SNR. Later, when the reverse shock of the SNR passes through the PWN its symmetric
shape is disturbed and after the reverberation phase the new smaller and often spherically
symmetric PWN can be formed. At this evolutionary stage, if the pulsar moves trough its SNR with a
supersonic speed the former spherical PWN takes on a cometary appearance. At this stage the
detailed MHD simulations provide significant information about the evolution of the PWNe
\cite[see][for the recent review of the MHD simulations]{Bucciantini13}.

In our model, the thermal absorption strongly depends on the orientation of the line-of-sight with
respect to the asymmetric bow-shock PWN.  The structure of such PWNe can be found e.g. in
\citet[][see also our Fig.3]{Bucciantini02}. There are three main regions of a PWN (see Fig.3): (1) the free
pulsar wind region, (2) the shocked pulsar wind region and (3) the shocked ISM that is interacting
with the shocked pulsar wind material. Since the strength of the free-free absorption depends on
$N_e^2$ we concentrate on the regions (2) and (3) as they can potentially have the highest
densities.

\begin{figure*}
   \resizebox{15cm}{!}{\includegraphics{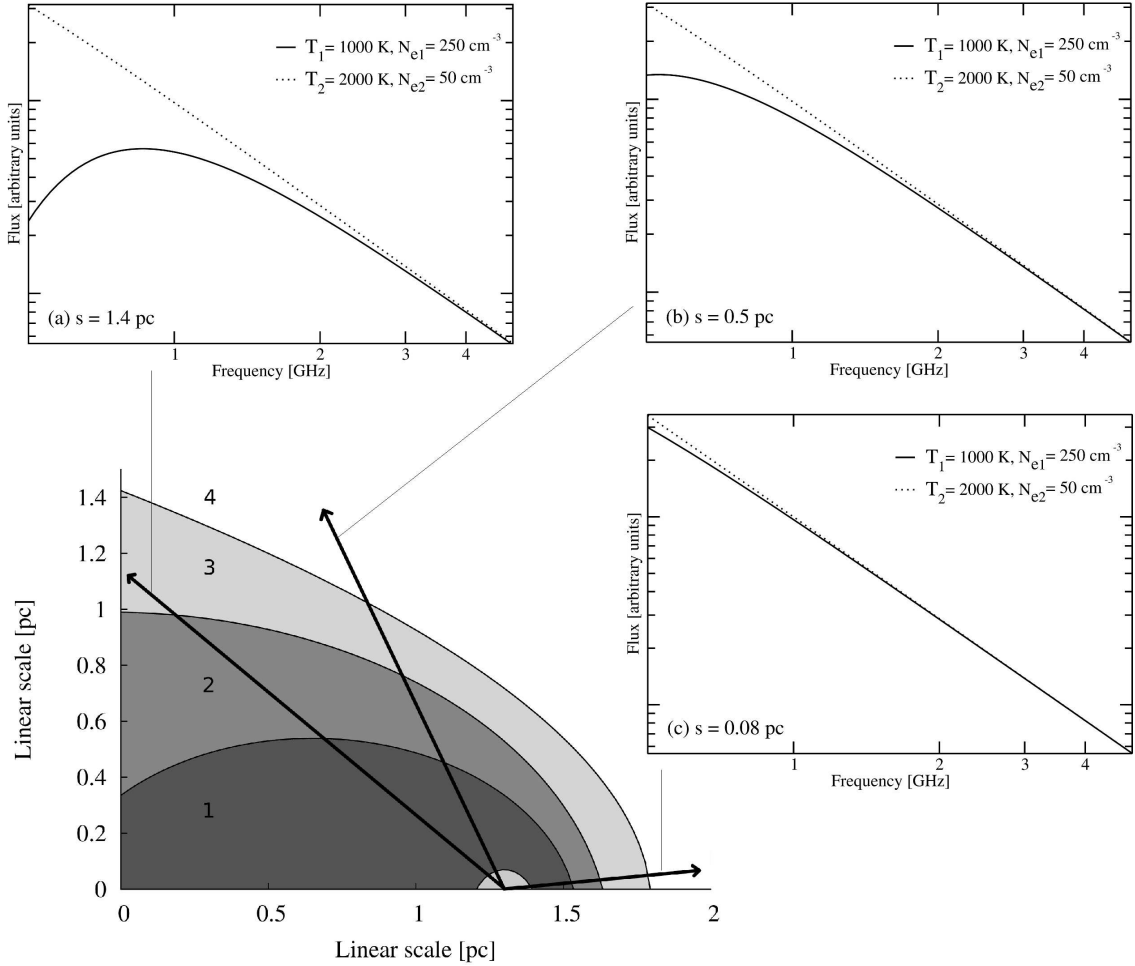}}
   \caption{
A schematic view of the bow-shock PWN (based on the MHD simulations \citealt{Bucciantini02}). The
numbered PWN regions are: (1) the free relativistic pulsar wind region, (2) the shocked pulsar wind
region, (3) the shocked ISM interacting with the shocked pulsar wind material, (4) - the ISM. The
termination shock is located between (1) and (2) and the forward shock - between (3) and (4). The
simulated apparent pulsar spectra for various orientations of the line-of-sight are presented in
the sub-panels; each of them shows two spectra obtained for different sets of physical parameters.
Obviously, in some parameter domains the same pulsar-PWN system can yield a different pulsar
spectrum depending on the direction of the line-of-sight (for detailed discussion see the
text).\label{fig3}}
\end{figure*}

Figure~\ref{fig3} presents the outcome of the free-free absorption in the cometary-shaped PWN. For
simplicity we have assumed both density and temperature profiles to be uniform over the regions
that contribute to the absorption. These are marked as (2) and (3) in the lower left part of
Figure~\ref{fig3} (the structure of the PWN is based directly on \citealt{Bucciantini02}). The
spectra plots shown in the separate panels present the possible shapes of the apparent pulsar
spectra (modified by the absorption) for two different sets of parameters, namely the temperature
$T$ and the electron density $N_e$ (the values used are presented in the plots). The panels (a),
(b) and (c) in Figure~\ref{fig3} show the apparent shape for three different orientations of the
line-of-sight corresponding to the observers location behind (a), at the side (b) and in front of
the bow-shock PWN (c). Since we assume the uniform density and temperature profiles, the thickness
of the absorbing region is the only parameter that varies along with the orientation of the
line-of-sight. This quantity can change from a fraction of a parsec (if an observer is in front of
the PWN (c)) up to well over a parsec (if an observer is behind the PWN (a)). This thickness
significantly influences the optical depth and hence the apparent spectrum of the pulsar that is
located in the middle of the PWN. In the case presented in Figure~\ref{fig3} panel (a) the spectrum
looks like a GPS with a peak frequency close to 1~GHz, while in the case presented in panel (b) the
spectrum can be interpreted (depending on the observational frequency range) as a pulsar with a
``broken spectrum''\footnote{Broken spectrum is a class of pulsar spectra which is often
morphologically described by two power-laws, with the lower frequency slope being flatter.}.
However, in the case presented in panel (c) the spectrum hardly differs from its intrinsic single
power-law spectrum.

The details of the absorption strongly depend on the actual physical parameters of the absorbing
electrons. For some sets of parameters there can be almost no absorption at all regardless of the
line-of-sight orientation (see the spectra presented by dotted lines in Fig.~\ref{fig3}).

The estimates of the electron density in the PWNe are quite rare. The only measurement we could
find in the literature is that of the PWN around PSR B1951+32, where \citet{Hester1989} and
\citet{Li05} estimated the electron density in front of the forward shock to be of the order of
$N_e = 50-100$ cm$^{-3}$ (from [O~III], [S~II] and [N~II] emission lines). However, it should be
mentioned that this value concerns the density of the uncompressed matter outside the shock while
near (or in) the PWN's bow-shock the density can be larger.

Of course we realise that our PWN model is quite simplified, and in reality the profiles of both
temperature and density may not be uniform. However, we believe that even such a simplified model
shows that in the case of the asymmetric PWN the pulsar apparent spectrum strongly depends on the
line-of-sight orientation and the influence of the nebula can be quite significant. Therefore, even
this model is able to explain why most of the GPS pulsars are found in the PWNe, while at the same
time a large fraction of pulsars in the PWNe does not show any kind of spectral deviation.

\begin{figure}
\includegraphics[angle=0,width=6.9cm]{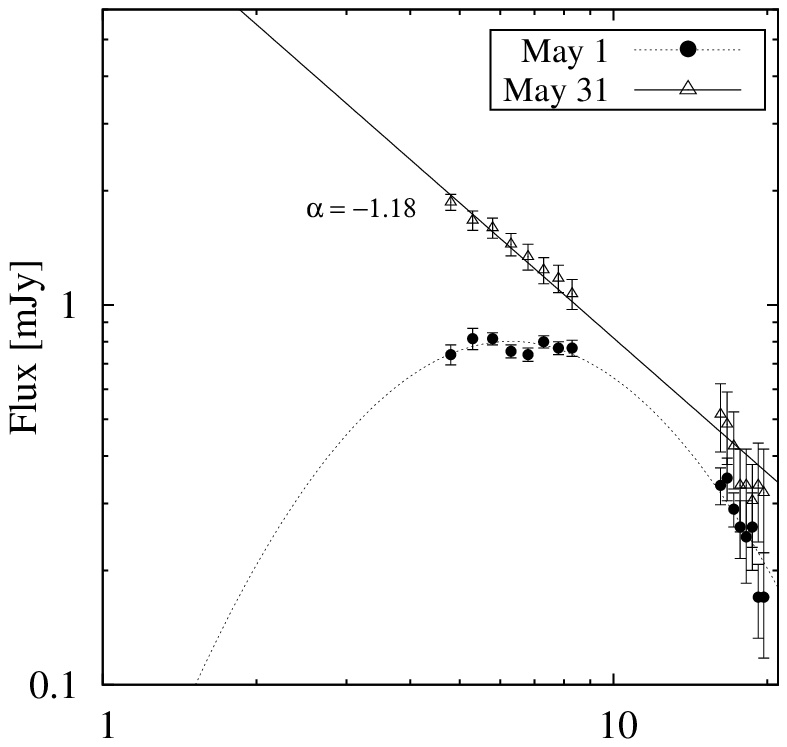}
\includegraphics[angle=0,width=6.9cm]{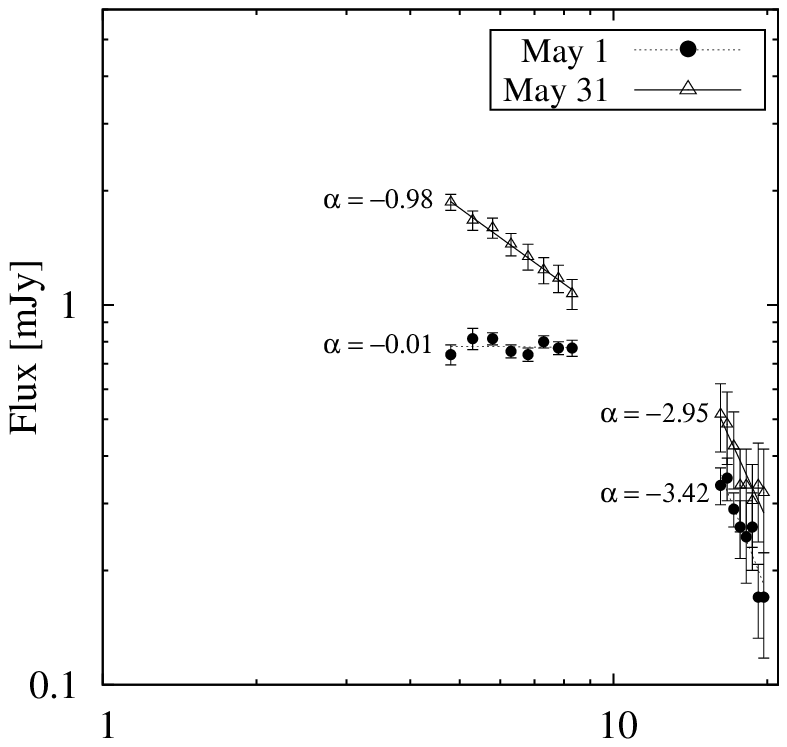}
\includegraphics[angle=0,width=6.9cm]{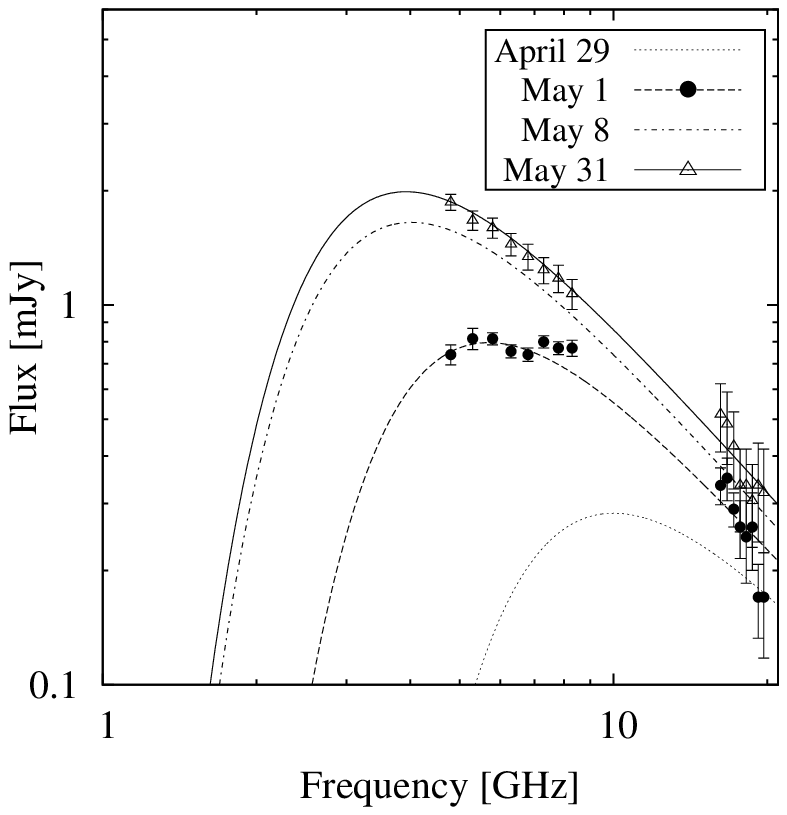}
\caption{
Spectra of the radio-magnetar SGR J1745-2900. Full dots denote the 2013 May 1 observations, empty
triangles - the 2013 May 31 measurements (all values are taken from Fig.4 in \citealt{Shannon13}).
The top panel shows our fits to the entire frequency range: the parabolic fit to the May~1st data
and a power-law fit for the May 31 observations. The middle panel shows two power-law fits (i.e. a
``broken spectra'' fits) that are performed separately for the two frequency bands used in the
observation. The bottom panel show the results of the absorption model fits for the observed data,
as well as the spectra modeled for April 29 and May 8 (see section 2.4 for details of the model).
\label{fig4}}
\end{figure}

\subsection{Sgr~A* radio-magnetar}

As we have mentioned in Introduction, \citet{Kijak13} demonstrated that the spectra of two
radio-magnetars seem to exhibit the GPS behavior with the peak frequency of a few gigahertz (5.0
and 8.3~GHz for SGRs J1550$-$5418 and J1622$-$4950, respectively). \citet{Kijak13} suggested that
the thermal absorption of radio waves is responsible for the GPS appearance in the case of the
radio-magnetars in the same way as it is in the case of the normal GPS pulsars\footnote{Let us
note, that a mechanism for radio emission of magnetars should not differ essentially from that of
the normal pulsars \citep{Szary15}.}. However, the reason for the very high peak frequencies is
still unclear. It should be mentioned that there is a lack of normal GPS pulsars with such high
peak frequencies. That may be explained by observational selection effects as it is very
difficult to find such objects in the usual pulsar search surveys, which are performed mostly at
frequencies at which the GPS-pulsars with a high peak frequency could be very faint.

Another, and probably the most famous (at least nowadays) radio-magnetar SGR J1745$-$2900, also
known as the Sagittarius~A* magnetar (due to its location at the center of the Milky Way) was
discovered by the \textit{Swift} Telescope on April 25, 2013 \citep{Kennea13}. The \textit{NuSTAR}
observation on April 26, 2013 found the object variability with the period $\sim 3.76$~s
\citep{Mori13}. These observations led to an identification of this source as a new magnetar
undergoing an outburst. Later Sgr A* magnetar was observed by radio-telescopes and consequently was
first detected at higher radio frequencies and later at frequencies as low as 1.5~GHz. The flux
density measurements (amongst other data) were performed by several observing groups using several
telescopes around the world: the Effelsberg radio telescope, the VLA, the telescopes at Nan\c{c}ay,
Jodrell Bank, Green Bank and Parkes
\citep[see][and references therein]{Eatough13, Rea13}.

Further observations made by the Australia Telescope Compact Array (ATCA) provided the data for the
magnetar radio spectrum for two epochs: May 1 and  May 31, 2013 and for two frequency ranges: from
4.5 to 8 GHz and from 16 to 20 GHz  \citep{Shannon13}. Figure~\ref{fig4} presents (in log-log
scale) the SGR J1745$-$2900 spectra obtained for these two epochs: the circles denote the
measurements from May 1, while the triangles denote May 31 data (values taken from Fig.4 by
\citealt{Shannon13}). The top panel in Figure~\ref{fig4} shows fits to the whole dataset that was
based on the observations in two frequency bands. The data from May 1 definitely do not resemble a
typical pulsar spectrum (a single power law) and thus suggest the possible GPS case. We have
performed a log-parabolic fit to the data (i.e. the fitted function was $S(\nu)=10^{(a
\log^2\nu+b\log\nu+c)}$, see \citealt{kuzmin2001}) and as a result we have found a possible maximum in the spectrum
at 6.3~GHz. This value seems to be similar to the peak frequencies that was found for other two
radio-magnetars mentioned above. We have attempted to fit a single power-law spectrum to the data
obtained on May~31 as a slight bend noticeable in the data is not big enough to indicate the
maximum of the spectrum. As a result we have obtained a spectral index of $\alpha=-1.18\pm0.04$ but
as one can see in the left panel of Figure~\ref{fig4} the power law (indicated by the dashed line)
fits the lower frequency band almost perfectly, while for the higher frequency band the fit is not
so good.

The reason why the fit at high frequency band deviates from the data points is the chosen method of
fitting, namely the so called weighted fitting method. As the uncertainties of the flux
measurements at the higher frequency were much larger than of that at the lower frequencies, the
fit weights of the high frequency data were much lower than that of the lower frequency data. On
the other hand, the distribution of the  16-20 GHz band data points (with respect to the fit line,
see the middle panel of Figure~\ref{fig4}) is not random and show an apparent trend indicating a
steeper slope. This is why we have performed the fitting separately for the lower and higher
frequency ATCA bands. The obtained fits are presented in the right panel of Fig.~\ref{fig4}. For
the data from May 1 we obtained a spectral slope of $-0.01\pm0.08$ for the 4.5-8.5 GHz band and
$-3.42\pm0.57$ for the 16 to 20 GHz range. Similar fits have been performed for the May~31 data
yielding $\alpha=-0.98\pm0.03$ at lower frequencies and $\alpha=-2.95\pm0.50$ at higher
frequencies. As one can see in Figure~\ref{fig4} the two spectral slopes (obtained by the data from
May 1 and May 31) in the higher band are almost the same for both epochs (the slops coincide within
the error bars), while in the lower frequency the spectra apparently differ from each other. It is
obvious, that the pulsar was significantly weaker during the first observation, but also the
spectrum on that date seems to be flat, while a month later it is definitely steeper.

We believe that this kind of behaviour may be caused by a thermal free-free absorption in the wind
nebula around Sgr A* radio-magnetar. On May 1, 2013 the optical depth of the nebula was higher,
hence the absorption was stronger, due to the higher electron density. Therefore, the
radio-magnetar exhibited the GPS behavior. However, a month later the optical depth decreased due
to dissipation of the absorbing material and consequently the absorption became weaker but probably
still observable on May 31, 2013.

Since the free-free absorption affects the higher observing frequencies to a much lower
extent, the slope in the higher frequency band stays unchanged. At the same time the flux at lower
frequencies gets stronger and in the 4.5 to 8.5 GHz band the spectra becomes sloped, which can
indicate that the absorption decreases. 

The explanation given above, as well as the idea of the spectral evolution in general, does not
contradict with the other observations using various radio-telescopes \citep[see,
e.g.][]{Eatough13, Rea13}. These observations, which were carried out over a wide range of
epochs, indicated that after few days of unsuccessful attempts the pulsar was first detected at the
higher frequencies. Additionally, despite the various attempts the first detection at 1.5 GHz
became possible only in mid-May, i.e. 2 to 3 weeks after the outburst from April~25. It should be
mentioned that the observations carried out on different dates cannot be used for obtaining the
spectra as it varies considerably over several days.

We can speculate on the cause of the decrease of the absorption, assuming that we deal with the
case of free-free absorption. We find it unlikely that the rate of absorption would change due
to the temperature variations. If it were so, the temperature would have to increase but there is
no apparent reason for a growth of the temperature. Another obvious way to decrease absorption is
to decrease the electron density, as the absorption strength is very sensitive to density
variations. We can suggest that the simplest possible explanation of the SGR J1745$-$2900 radio
spectra evolution is the absorption in the electron material ejected during the magnetar outburst.

Initially, the density of absorbing material was so high that even at the higher observing
frequencies the radio pulsations were undetectable.  If one assumes that the ejected material
expands in a form of a spherical shell and with a constant velocity, the density decreases with
distance as $1/r^2$ (or with time as $1/t^2$) the optical depth of the absorbing medium falls even
faster than the density ($1/r^4$). Therefore, it was possible to detect the radio-magnetar just in
a few days after the outburst at higher radio frequencies, while at lower frequencies detection
became possible only in few weeks. One can argue that the decrease of the optical depth due to the
decreasing density might, at least to some extent, be compensated by the decrease of the
temperature of the ejecta. However, the optical depth is much more sensitive to the variation of
the density than to that of the temperature (see Eq.~\ref{opt_dep}). In addition we have assumed
that the cooling of the matter is an adiabatic process (as it is usually assumed in the cases of
outbursts and explosions, for example in the supernovae models) which usually leads to a linear
drop in temperature with distance ($T\sim 1/r$). Thus, the net effect should be a decrease of the
optical depth that leads to the decrease of absorption and to the shift of the peak frequency
toward lower frequencies.

In the case of PSR~J1745$-$2900 it is impossible to develop a detailed model based only on the
observational data. To explain the amount of absorption responsible for the appearance of the
spectra one needs information about the electron density, temperature and thickness of the
absorber. The only parameter that could be associated with the observational data is the
initial temperature of the ejecta which in our model is assumed to be the same as the temperature
inferred from the X-ray spectrum of the outburst. As for the detailed model of the relation between
these temperature(s) it is still under debate. These data, based on the X-ray spectrum observed
during the outburst \cite[see][]{Kennea13} indicate two different temperatures of the ejecta: the
first one, $10^8$~K, was observed on April~25 during the discovery observation of the outburst and
the second, lower one, $10^7$~K, was obtained based on the integrated data obtained during very
long observation sessions after the outburst. In our model we used the latter value. In our model
the temperature of the ejecta decreases with time linearly. However, it cannot decrease
indefinitely (if that were the case then on May~1, just 6 days after the outburst it would drop to
100~K). One has to remember that this magnetar is located in a densely populated region of the
Galactic center and the electron gas should be always heated by the background radiation, just as
it happens in a typical HII region. For typical ionized regions in the galactic center the electron
temperature is in the range from 5000~K to 7000~K \citep{afflerbach1996}. Therefore, we have
adopted 5000~K as the lowest possible temperature of the ejecta, i.e. in our model the temperature
never drops below this value. As for the other parameters that influence absorption we cannot
decouple the value of the electron density from the thickness of the expanding spherical layer,
since we do not have any observational data for these parameters. Hence, we have decided to assume
the initial value of the electron density and fit the model to the observational data by adjusting
the thickness of the absorber.

While trying to model both the May~1 and May~31 spectra we have come to the conclusion that any
absorption that enables us to explain the shape of the May~1 spectrum will completely disappear in
31 days. Hence, in order to explain the shape of the May~31 spectrum we have introduced constant
absorption in addition to the expanding ejecta. Our initial assumption was that such an external absorption 
may be provided by either the interstellar medium or/and some kind of a residual PWN around the 
radio-magnetar, which may exist due to the previous  outbursts of the source\footnote{The possibility of the existence of such a
PWN around the Sgr~A* magnetar was recently discussed by \citet{Tong14} who also stated that the
PWN would be undetectable for the current telescopes due to an extremely large distance to the
source which is located in the Galactic Center.} However, as we discuss below, this is rather not the case.

Since we have no information about the physical parameters of this external absorber we have
decided to explore two possibilities. One possibility is to assume that the absorption happens in a
cold medium, such as a dense (and partially ionized) molecular cloud, which is similar to the SNR
filament we discussed above. If we assume the temperature of a cloud to be 200~K and its thickness
to be about 1~pc, then the electron density in the cloud should be close to 500~cm$^{-3}$ in order
to satisfy the May 31 absorption. On the other hand, if we assume that there is no such cloud in
the magnetar's line-of-sight and the absorption happens in a hot medium (5000~K) then in order to
satisfy the observed absorption the thickness of the absorbing medium should be about 0.15 pc and
the electron density should be about $10^4$~cm$^{-3}$. Let us note, that this value is a typical
for the Galactic Center HII regions, see \citealt{genzel95}. The fit (which is obviously the same
for both of these sets of parameters) to the spectrum is shown in the bottom panel of
Fig.~\ref{fig4} by the solid line. It is worth mentioning that to obtain this specific spectrum one
can use a lot of sets of parameters and the set mentioned above represents only some of the
possibilities. The validity of these parameter sets is discussed later in this Section based on
other observable properties of the magnetar.

In our modeling of the May~1 spectrum the temperature of the ejecta is already the lowest allowable
value, 5000~K. Thus we have to use some reasonable values for the initial electron density and the
geometrical thickness of the shell. Taking into account the external absorber we can obtain the
shape of the May~1 spectrum if we assume that the electron density is $N_e\simeq2\times
10^5$~cm$^{-3}$ and the geometrical thickness of the shell is close to 0.7 light-days (or $5\times
10^{-4}$~pc). These values seem quite reasonable for the May~1 (6 days after the initial outburst)
conditions. The value $N_e=2\times 10^5$~cm$^{-3}$ requires the initial (i.e. at the stellar
surface) electron density to be equal to $5\times10^4$ of the Goldreich-Julian number density which
is estimated as $3\times10^{12}$~cm$^{-3}$ at the surface of this radio-magnetar. The used value
for the thickness of the ejecta can be reached if the thermal spread of the electrons in the ejecta
is about 10\% of the average speed of particles. However, it should be remembered, that this
thickness is closely bound to the assumed initial density, and if the density is lower one will
need to increase the thickness of the absorber significantly in order to explain the observed
effect.

The resulting spectrum is shown in the bottom panel of Fig.~\ref{fig4}. Let us also note that to
satisfy the shape of the spectrum we have to assume that the intrinsic radio spectrum of the
magnetar has a spectral index $-1.5$. This value is significantly lower than the one suggested by
the observational data from the higher frequency band (compare the middle panel of Fig~\ref{fig4}).
However, the use of the steeper intrinsic spectrum results in a much sharper peak in the spectrum
after applying the absorption. Thus, it makes our model unable to explain the almost flat spectrum
in the lower frequency band of the May~1 data. It should be mentioned that the physical parameters
of the absorbing ejecta quoted above represent only one of the possible combinations that can
explain the observed absorption.

After fitting our model to the data we have used this model to estimate the shape of the Sgr~A*
magnetar spectra for other epochs, for the purpose of showing the spectrum evolution. In the bottom
panel of Fig.~\ref{fig4} besides the curves fitted to the data we have also presented the modeled
spectra for April~29 (4 days after the initial outburst) and May~8 (two weeks later). This
evolution can easily explain why the radio magnetar was not detectable at lower radio frequencies
(i.e. from 1 to 3 GHz range) for almost two weeks after the outburst.

It is worth mentioning that as a by-product of our model we have ability to estimate the
contribution made by the absorber to the magnetar's dispersion measure. The DM of PSR~J1745$-$2900
was reported to be in the range between 1650 and 1830~pc~cm$^{-3}$ (various authors quote different
values, see \citealt{Eatough13, Rea13, Shannon13}). Assuming the homogenous density distribution
and using the expression for the DM, $\Delta DM = N_e d$ (here $d$ is the thickness of the absorber
in parsecs), we can estimate the amount of DM added by the ejecta to the magnetar's DM for the
May~1 conditions. This additional value turns to be of about 100~pc~cm$^{-3}$ and obviously
drops with time due to the decrease in the electron density. Thus, this value is smaller than
the discrepancy between the DM values given by various authors. Additionally the published DMs are
usually given without the exact epoch of the measurement.

We can also estimate the contribution to the total DM from the material that provides what we call
the external absorption. If we assume that this absorption happens in the cold molecular cloud
(with the thickness of $1$ pc, $N_e=500$~cm$^{-3}$ and $T=200$~K) which is one of the possibilities
suggested earlier, then its contribution to the total DM can be estimated as 500~pc~cm$^{-3}$,
which is significantly greater than that of the ejecta.  We believe that this can explain the
observed value of the DM as compared to the DMs of other pulsars that lie close to or behind
the Galactic Center. The DMs of these pulsars rarely exceed 1200~pc~cm$^{-3}$.

On the other hand we know that the magnetar is located extremely close to the Galactic Center (0.09
pc) and the surrounding ISM resembles a hot HII region (the second set of parameters we have
discussed earlier: the thickness of $0.15$ pc, $T=5000$~K and $N_e=10^4$~cm$^{-3}$). Therefore, its
contribution to the total DM would be of the order of 1500~pc~cm$^{-3}$, which is clearly too high.
Since the above-mentioned parameters of the absorbing region are only tentative, we can freely
adjust them. Let us note that the optical thickness $\tau \propto d \ N_e^2$ and the contribution
to the total DM is equal to $d N_e$. Thus, higher electron densities and smaller sizes are needed
in order to provide the same observed absorption, but with a smaller contribution to the total DM, i.e. an
absorber with $N_e=1.8 \times 10^4$~cm$^{-3}$ and the thickness of 0.05 pc would contribute only
900~pc~cm$^{-3}$ to the total DM, assuming the same temperature of $5000$~K. The further decrease of
the DM contribution requires really small sizes of the absorber region which seems to be rather
unlikely. Therefore we believe that a cold ISM cloud is the most likely candidate for an absorber
in this case.

Obviously our calculations of the contributions to the magnetar's
DM provided by both the ejecta and the external absorber are based on specific density and thickness estimates
that follow from our model. Since a number of different sets of parameters can satisfy the observed
absorption and the resulting DM contribution is variable, these calculations should be considered
as rough estimates.

To summarize our discussion on the Sgr~A* radio magnetar we can explain, at least to some extent,
the basic properties of the  observed evolution of its radio spectrum by the free-free absorption
in the expanding electron cloud that was ejected during the outburst. However, our ability to
perform a detailed study is limited due to the sparseness of observations (i.e. multi-frequency
flux density measurements), as we have somewhat reasonable spectra for two epochs only.
Nevertheless, we believe that our model provides an adequate explanation of the observational data.
Also, the PSR~J1745$-$2900 radio magnetar is so far the only one for which we have relatively
detailed spectra for certain epochs (thanks to the data from \citealt{Shannon13}). Unfortunately we
cannot apply our  model to the spectra of the other two GPS magnetars since their spectral data are
much more sparse. We can only hope that in the future observations of the radio-magnetar outbursts
the assessment of the evolution of the spectrum and the DM variations, especially within the first
several days, will become one of the priorities.

\section{Conclusions}

In this paper we have explored the possibility that the gigahertz-peaked spectra (observed in a
number of pulsars and radio-magnetars) may be caused by the thermal free-free absorption in the
electron mater of the neutron star surroundings \citep[see also ][]{Kijak11b,Kijak13}. Using the simplified model, which assumes a
constant density and temperature profiles in the absorbing media, we were able to simulate the
spectra of pulsars that resemble the spectra of the known GPS objects at least qualitatively. We
have discussed several possible scenarios and geometries of the pulsar surroundings that can
provide the absorption that makes a pulsar spectrum peak at the frequencies 1 GHz and above. One of
these scenarios is the case of very dense filaments in the supernova remnants, and we have shown
that at least some of the observed filaments are dense and possibly cold enough to affect the
pulsar spectrum in a desired way. The fact that such kind of absorption indeed requires the most
dense (and thus obviously relatively small) parts of an SNR to be involved may explain why not all
pulsars associated with SNR show the GPS signature in their spectrum.

Another possibility we have explored is the thermal absorption in the pulsar wind nebulae, and our
simulations show that these objects may indeed be a cause of the GPS if the geometry is favorable,
i.e. the same kind of PWN depending on the orientation of the line-of-sight may or may not provide
enough absorption to cause the pulsar spectrum to peak at a few GHz. We have found that in order to
observe the significant absorption in the cometary-shaped bow-shock PWN the observer should be
located behind the PWN (see panel (a) of Figure~\ref{fig3}). At the same time, this geometrical
dependence may explain why some of the PWN associated pulsars do not exhibit a GPS phenomenon.
Although beyond the geometry the free-free absorption may also strongly depend on the physical
properties of a PWN, and hence also on its evolutionary phase.

The other two cases that we have studied include the spectral evolution. In the case of the binary
system PSR~B1259$-$63/LS2883 (discussed in detail by \citealt{Dembska14b}) the evolution occurs due
to the pulsar orbital motion that changes the orientation of the line-of-sight with regard to the
absorbing region and, therefore, the absorption rate increases significantly when the pulsar moves
deeper in the dense stellar wind of its Be-star companion. In the case of the Sgr~A* radio-magnetar
(which would be the 3rd magnetar with a GPS) the spectral evolution is caused by the free-free
absorption of the magnetar radio emission in the electron material ejected during the outburst. The
ejecta expands with time and consequently the absorption rate decreases and the shape of the
spectrum changes in such a way that the peak frequency shifts towards the lower radio frequencies.

\section*{Acknowledgments}
This work was supported by the grants DEC-2012/05/B/ST9/03924 and DEC-2013/09/B/ST9/02177 of the Polish National Science Centre. We are grateful to the anonymous referee for many constructive comments that helped us to significantly improve this paper.

\label{lastpage}
\end{document}